\documentstyle[preprint,aps,prl,epsfig]{revtex}
\begin{document}
\draft
\narrowtext
\tighten

\title{Probing the $^{6}$He halo structure with elastic and inelastic
proton scattering}
\author{A.~Lagoyannis$^{1,2}$, F.~Auger$^1$,
A.~Musumarra$^1$\footnote{present address:INFN-Laboratori Nazionali del Sud, 
Via S. Sofia 44, 95123 Catania Italy}, N.~Alamanos$^1$, E.~C.~Pollacco$^1$,
A.~Pakou$^2$,
Y.~Blumenfeld$^3$, F.~Braga$^1$, M.~La~Commara$^4$, A.~Drouart$^1$,
G.~Fioni$^1$, A.~Gillibert$^1$, E.~Khan$^3$, V.~Lapoux$^1$, W.~Mittig$^5$, 
S.~Ottini-Hustache$^1$, 
D.~Pierroutsakou$^4$, M. Romoli$^4$,
P.~Roussel-Chomaz$^5$, M. Sandoli$^4$, D. Santonocito$^{3,*}$,
J.~A.~Scarpaci$^3$, J.L.~Sida$^1$ and T.~Suomij\"arvi$^3$}
\address{$^1$DSM/DAPNIA CEA SACLAY, 91191 Gif-sur-Yvette, France}
\address{$^2$Department of Physics, The University of Ioannina, 45110 Ioannina,
Greece}
\address{$^3$Institut de Physique Nucl\'eaire, IN2P3-CNRS,
F-91406, Orsay, France} \address{$^4$University of Napoli and INFN
Sezione di Napoli, I-80125, Napoli, Italy} \address{$^5$GANIL, BP
5027, F-14021, Caen, France}
\vspace{5mm}
\author{S.~Karataglidis$^{6,7}$ and K.~Amos$^{8}$}
\address{$^6$TRIUMF, 4004 Wesbrook Mall, Vancouver, British Columbia,
V6T 2A3, Canada}
\address{$^7$Theory Division, Los Alamos National Laboratory, Los
Alamos, New Mexico, 87545}
\address{$^8$School of Physics, University of Melbourne, Parkville
3052, Victoria, Australia}
\date{\today}
\maketitle

\begin{abstract}
Proton elastic scattering and inelastic scattering to the first
excited state of $^6$He have been measured over a wide angular range
using a 40.9$A$ MeV $^6$He beam. The data have been analyzed with a
fully microscopic model of proton-nucleus scattering using $^6$He wave
functions generated from large space shell model calculations. The
inelastic scattering data show a remarkable sensitivity to the halo
structure of $^6$He.
\end{abstract}

It is well known that neutron rich weakly bound light nuclei have
abnormally large radii\cite{Ta88}. This phenomenon is attributed to
the valence neutrons which tunnel out of the core so that they have a
large probability to be present at distances greater than the normal
nuclear radius. Considerable experimental and theoretical efforts have
been devoted to the understanding of the structure of these so-called
halo nuclei\cite{Ha95,Jon97}.  However, due to the low intensities of
the available exotic beams, it is only recently that inelastic
scattering and transfer reactions on light particles, which are the
best tools to probe deeply the structure of nuclei could be undertaken
under good conditions.  Experimentally, the Borromean $^6$He nucleus
is the best candidate for this kind of study since the only confirmed
discrete state is a 2$^+$ state situated at 1.87~MeV \cite{Aj88}. It
has been investigated through the measurement of interaction,
dissociation and elastic scattering cross sections
\cite{expset}. However, apart from elastic scattering, those reactions
involve breakup of the $^6$He into its constituents ($^4$He $+n+n$),
and that has been shown to be significantly influenced by final state
interactions \cite{Al98}, which destroys information about the $^6$He
ground state.

To study the microscopic structure of $^6$He we measured elastic and
inelastic scattering of $^6$He from protons by making use of a new
large acceptance detector array MUST\cite{Blu99}. In this letter,
extended angular distributions are presented along with a fully
microscopic analysis using wave functions generated from large space
shell model calculations allowing all the nucleons of $^6$He to be
active in the field.  The inelastic scattering angular distribution to
the 2$^+$ state can only be reproduced if wave functions including a
neutron halo are used, demonstrating for the first time a strong
sensitivity of inelastic scattering data to the halo structure.
  
\begin{figure}
\epsfig{figure=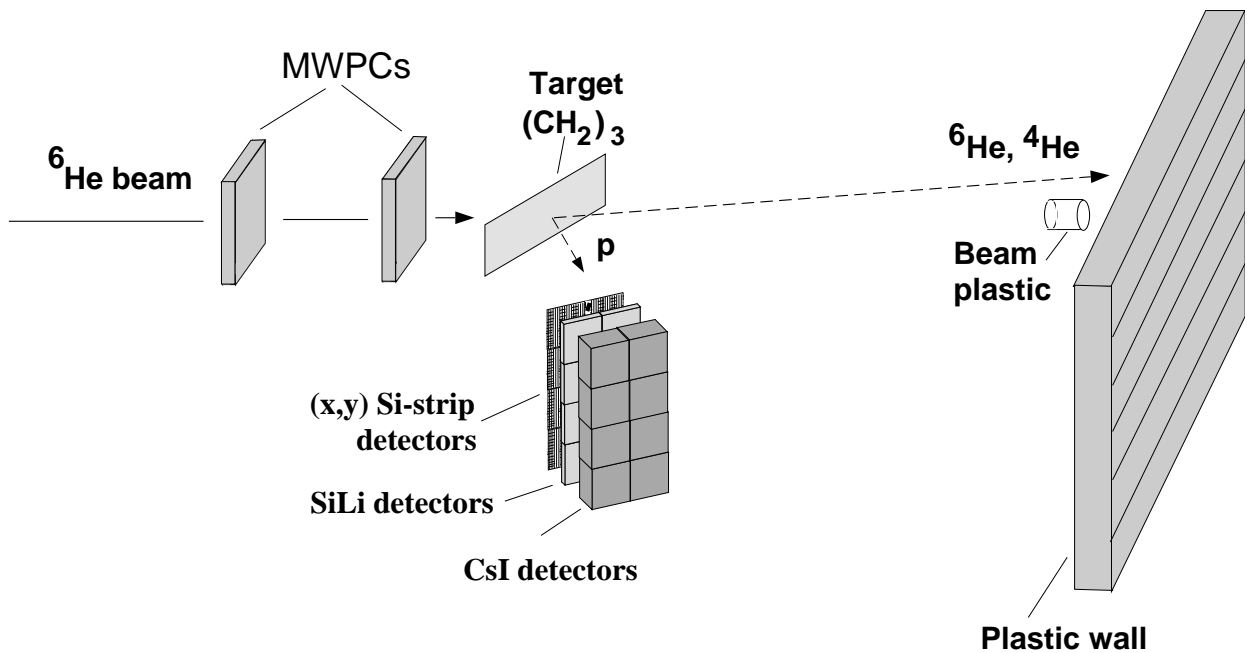,width=\linewidth,clip=}
\caption[]{The experimental set up.}
\label{setup}
\end{figure}

The experiment was performed at the GANIL facility with a $40.9A$~MeV
$^6$He radioactive ion beam produced by fragmentation of a primary
$75A$~MeV $^{13}$C beam on a 8.45~mm thick C target located in the
SISSI device\cite{Mi90}. The secondary beam was purified with a 0.9~mm
thick Al degrader situated between the dipoles of the
$\alpha$-spectrometer. The beam intensity on the polypropylene
(CH$_{2}$)$_{3}$ reaction target was $10^5$ particles per second with
a 2\% total contamination of $^8$Li and $^9$Be. As the beam spot on
the target covered 1~cm$^{2}$ with a maximum angular divergence of
$1^{\circ}$, two X and Y position sensitive detectors,
CATS~\cite{ott99}, were placed at 155~cm and 27~cm in front of the
target as illustrated in Fig.~\ref{setup}.  These detectors provided
the impact point and the incident angle on the target event by event
with a FWHM resolution of 0.55~mm (X), 0.7~mm (Y) and $0.1^{\circ}$.

The recoiling protons were detected in MUST~\cite{Blu99}, an array of
8 three-stage telescopes $6\mbox{ cm} \times 6 \mbox{ cm}$ each.  The
first stages consist in double-sided Si-strip detectors (300
$\mu$m). They were placed at 15~cm from the target and covered the
angular range between 46 and 90$^{\circ}$ in the laboratory frame. At
this distance, the 1~mm wide strips result in an angular resolution of
$0.4^{\circ}$ in both X and Y directions.  Protons of less than 6~MeV
were stopped in these detectors and were identified down to 0.5 MeV by
measurement of energy versus time of flight (E-TOF). The start of the
TOF measurement was given by the passage of the incident particle in
one CATS tracking detector and the overall time resolution was 1.2~ns.
Protons in the energy range of 6 to 25~MeV were stopped in the second
SiLi stage (3 mm) of the telescopes while those in the energy range
from 25 to 70 MeV were stopped in the third CsI stage (15 mm).  They
were identified by the $\Delta E- E$ method.

The ejectile was detected in coincidence with the recoiling proton to
suppress the protons emitted from excited nuclei produced in central
collisions of $^6$He on the carbon contained in the target. The
coincidence allowed also to suppress the protons coming from reactions
induced by the beam contaminants on the target.  The ejectile was
detected in a plastic wall, situated at 75~cm behind the target and
made up of 6 horizontal bars of BC408, $8 \times 50$~cm$^{2}$ and 3~cm
thick. Each bar was read out by a photo-multiplier at each end. The
large angular coverage of the wall was imposed by the in flight decay
of $^{6}$He ($^{6}\mbox{He} \rightarrow \alpha +2n$) that occurs for
excitation energies higher than the $2n$ separation energy of 0.9~MeV
\cite{Aj88}. Identification and counting of the beam particles were
achieved in a plastic scintillator with a diameter of 2.8~cm and
centered at zero degrees.

\begin{figure}
\epsfig{file=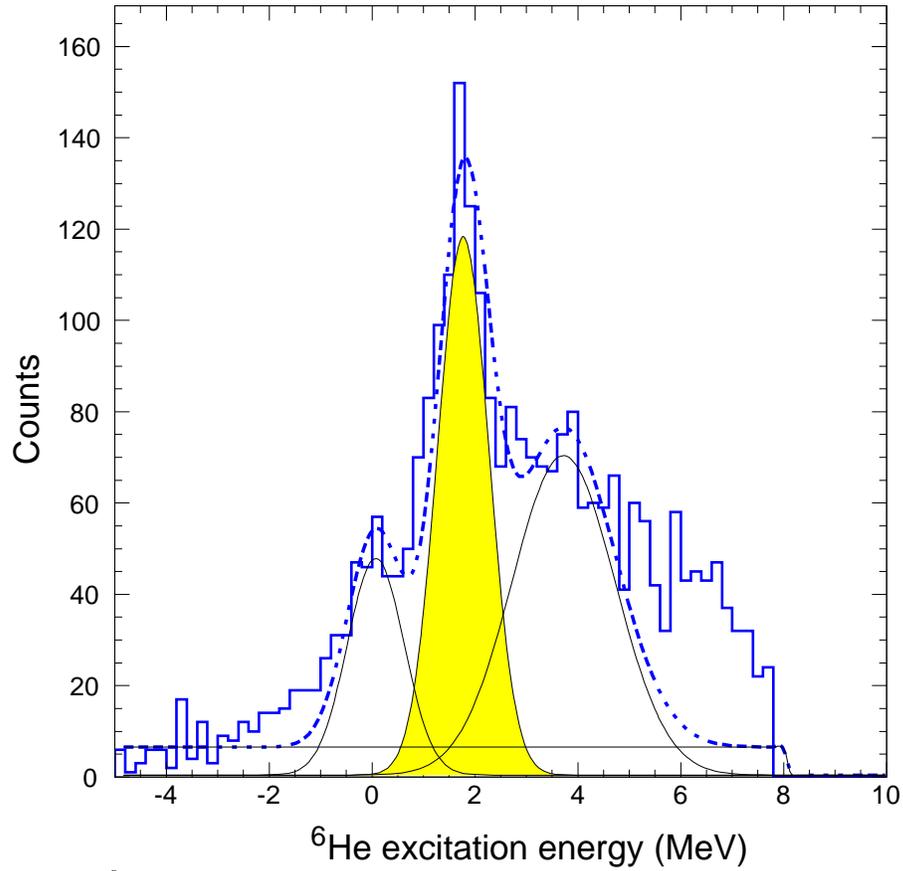,width=0.8\linewidth,clip=}
\caption[]{$^6$He excitation energy spectrum extracted
from protons measured between 60 and 70$^{\circ}$ in the laboratory in coincidence
with an alpha particle. Four components have been considered to estimate the 
background under the 2$^+$ peak (see text).} 
\label{spectrum}
\end{figure}

To measure angular distributions down to 10$^{\circ}$ in the center of
mass (85$^{\circ}$ in the laboratory) where the energy of the
recoiling protons decreases down to 0.5MeV, a 1.48 mg/cm$^{2}$ thick
polypropylene target was used. Good statistics at larger angles was
obtained by using a 8.25 mg/cm$^{2}$ thick target. Elastic
(respectively inelastic) events were extracted by requiring that a
proton be detected in coincidence with a $^6$He ( respectively alpha
particle).  Protons were selected with contours on the E-TOF and
$\Delta E- E$ planes of MUST and $^6$He (alpha particles) with
contours on the E-TOF matrix of the plastics. Events corresponding to
the excitation of the 2$^+$ state were extracted from inelastic events
by taking a window between 0.8 and 2.3 MeV on the excitation energy
spectrum of $^6$He calculated from the measured proton energy and
angle. As an illustration, the excitation energy spectrum of inelastic
events measured between 60 and 70$^{\circ}$ in the laboratory is shown
in Fig.~\ref{spectrum}. A small contamination remains from elastic
scattering and the high energy side of the 2$^+$ peak is contaminated
by low lying excitations in the continuum and by the fragmentation
processes.  In order to estimate the background under the 2$^+$ peak,
for each bin of 2$^{\circ}$ in the laboratory, the spectrum was fitted
with four components as shown in Fig.~\ref{spectrum}: a small constant
background corresponding to the background observed at the left of the
elastic peak; two Gaussians for the elastic and inelastic peaks having
the same width as the $^6$He elastic peak; and a third Gaussian on the
high energy side to simulate the excitations in the continuum. In
order to appreciate the uncertainty on the background subtraction, the
fit was also done with a linear component beginning at the 2n
separation energy (0.9MeV) instead of the third Gaussian. The
background represented between 10 and 30\% of the peak depending on
the angle with an uncertainty of $\pm$5\% on this percentage.  Elastic
and 2$^+$ state contributions were extracted for each 1$^{\circ}$ bin
in the laboratory frame and normalized with the acceptance of the
detection system, the target thickness ($\pm$5\% uncertainty) and the
number of incident $^6$He ($\pm$3\% uncertainty).

\begin{figure}
\epsfig{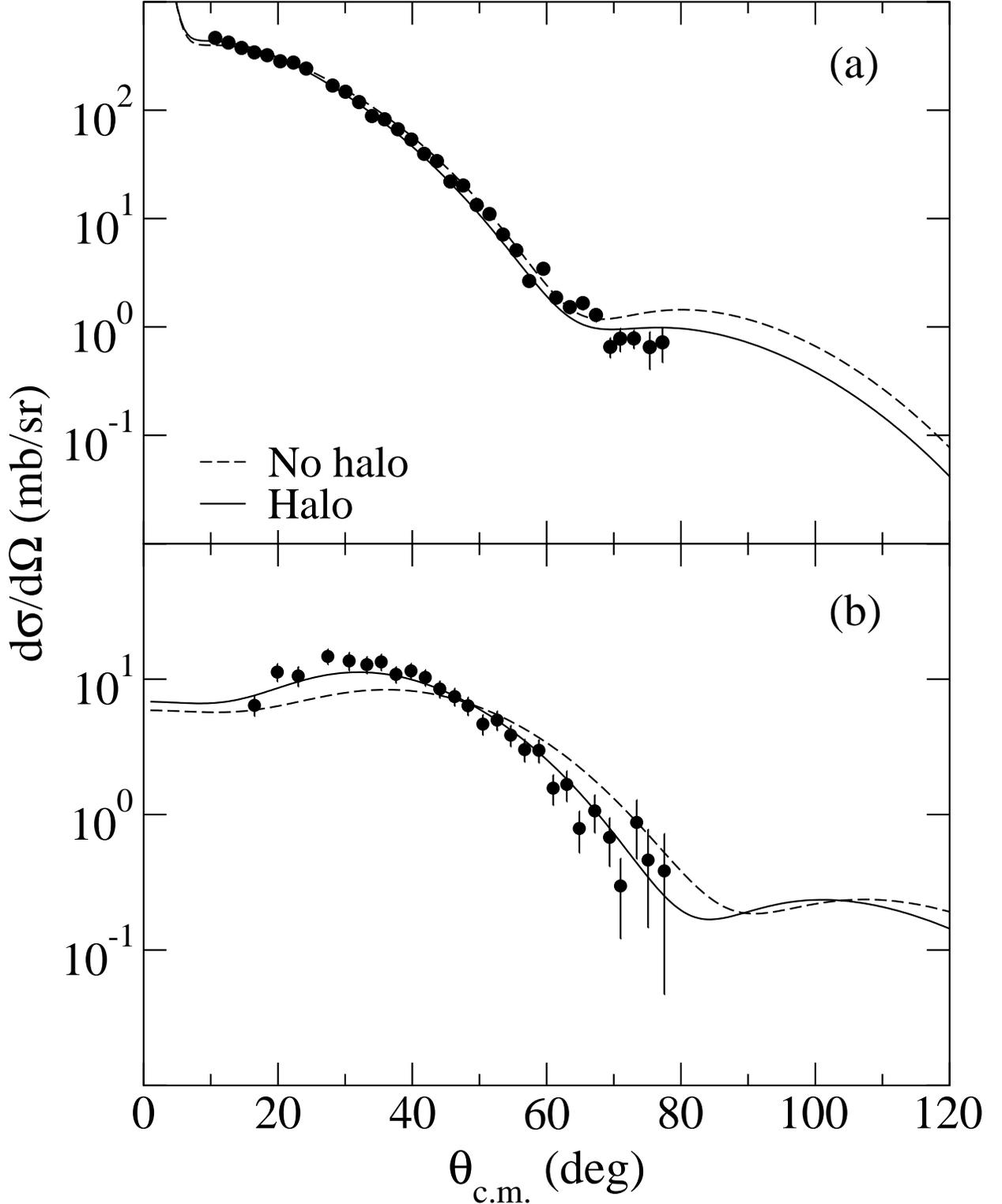}
\caption[]{Differential cross sections for the (a) elastic and (b) inelastic 
scattering to the of 2$^+$ state at 1.87 MeV of $^6$He
from hydrogen at $40.9A$~MeV. The present data (circles) are compared
to the results of the calculations assuming no halo (dashed line) and
halo (solid line) conditions.}
\label{results}
\end{figure}

Angular distributions in the center of mass are presented in
Fig.~\ref{results}. The error bars given for elastic scattering are
purely statistical whereas the error bars quoted for the inelastic
scattering include in addition the error due to the background
subtraction.

Calculations for the elastic proton scattering data were made using a
fully microscopic model of the optical potential~\cite{Do94}. In this
model, the potential is obtained in coordinate-space by folding a
complex energy- and density-dependent effective nucleon-nucleon ($NN$)
interaction with the one-body density-matrix elements (OBDME) and
single particle bound states of the target generated by shell model
calculations. As the approach accounts for the exchange terms in the
scattering process the resulting complex optical potential is
non-local.  This model has been applied successfully to calculate
elastic and inelastic scattering of protons from many stable and
unstable nuclei ranging from $^3$He to $^{238}$U at different energies
between 65~MeV and 300~MeV \cite{set,Ka97,Do97,Ka97a}.  The effective
interaction and the structure details were all preset and no {\it a
posteriori} adjustment or simplifying approximation was made to the
folded optical potentials. Hence the observables obtained are
predictions. Hereafter, we refer to the process of making such
potentials as $g$ folding.

Calculations of the transition amplitudes for the inelastic scattering
have been done within the distorted wave approximation (DWA).  The
same effective $NN$ interaction and shell model calculations used to
make the $g$ folding optical potential have been respectively used for
the transition operator and the transition OBDME. For the stable
nuclei whose spectroscopy is well defined from the measurement of
inelastic electron scattering form factor, the inelastic scattering
has been shown to be very sensitive, more than elastic scattering, to
the details of the effective interaction ~\cite{set}. Conversely, when
the effective interaction was well established, the analysis of
inelastic data turned out to be a very sensitive test of the model
structure used for the nucleus~\cite{set,Ka97}. As for elastic
scattering the calculations were parameter free.

To apply these models to 40~MeV proton scattering, the effective $NN$
interaction had to be determined. As for the higher energies, it has
been parametrized as a sum of central, two-body spin orbit and tensor
components, each of them being a set of Yukawa functions of various
ranges. This specific form is dictated by the structure chosen in the
program DWBA98~\cite{Ra98} which has been used for the analysis of
both the elastic and inelastic scattering data.  The complex, energy-
and density-dependent strength and the range of each Yukawa function
were obtained by accurately mapping the on- and half-on-shell $g$
matrices which are solutions of the Brueckner-Bethe-Goldstone
equations of the Bonn-B\cite{Ma87} realistic $NN$ interaction. The
validity of the 40~MeV effective interaction has been verified by
calculations of cross sections and analyzing powers of proton elastic
scattering for different stable nuclei~\cite{40MeVref}.

With the effective $NN$ interaction set, it remained only to define
the structure of $^6$He. A view that this nucleus should resemble an
$\alpha$ particle with two extraneous neutrons has fostered a
semi-microscopic cluster model treatment of the system~\cite{Ar99}.
On the other hand, large space (no-core) shell model
calculations~\cite{Na96,Ka00} and quantum Monte Carlo
calculations~\cite{Pu97} which are fully microscopic have been done.
The Navr\'atil and Barrett~\cite{Na96} large space shell model
calculations are suited to our scattering analyses.  They allowed the
6 nucleons to be active and their shell model interaction was
specified as $NN$ $G$ matrix elements~\cite{Zh95} generated from the
realistic CD-Bonn $NN$ interaction.  We used their complete
$6\hbar\omega$ wave functions to specify the relevant ground state and
$0^+ \to 2^+$ transition OBDME for $^6$He. To investigate the
sensitivity of the analyses on the size of the model space we have
also used wave functions from a complete $4\hbar\omega$ shell model
\cite{Ka00}. However, in both models the binding energy of the last
neutron is larger than the experimental separation energy 1.87~MeV
\cite{Aj88}. That would indicate that the size of the model spaces
used is still too small to give the correct asymptotic behavior of the
neutron density.

The $p-^6$He $g$ folding optical potential made with the shell model
prescribed HO functions is almost phase shift equivalent to that
obtained using WS functions which allow to fit electron scattering for
form factor of $^6$Li~\cite{Ka97}. This led us to use these WS
functions but in order to specify the neutron halo in $^6$He we
changed the bound state WS potential so that the 0p-shell binding
energy became 2~MeV which is close to the single neutron separation
energy.  Also the binding energies of the higher orbits were all set
to 0.5~MeV as more exact (smaller) values will not influence
noticeably results of the scattering.  The optical potential obtained
using these adjusted WS single particle wave functions leads to the
cross section hereafter designated as {\em halo}. The use of the HO
single particle wave functions given by either shell model leads to
the cross section that is designated as {\em no halo}. Use of the WS
wave functions has been shown to reproduce the r.m.s. radius of $^6$He
\cite{Ka00}.

The elastic scattering data are compared in Fig.~\ref{results}(a) to
the halo (solid line) and no halo (dashed line) calculations. The two
calculations are very similar up to 60$^{\circ}$ and notably differ at
larger angles. The agreement of the calculations with the data is very
good up to 60$^{\circ}$. The few data beyond these angles are better
reproduced by the halo description but it is clear that data at larger
momentum transfers are required to use elastic scattering as a probe
of the halo structure of $^6$He.

The very good agreement obtained with the elastic scattering data is
essential since it validates the $g$ folding optical potential used to
define the distorted waves in the DWA analysis of the inelastic
scattering leading to the $2^+;T=1$ state. Halo (solid line) and no
halo (dashed line) calculated cross sections for the 2$^+$ state are
presented in Fig.~\ref{results}(b).  Contrary to the elastic
scattering, the sensitivity to the halo is important over the entire
angular domain. The data are very well reproduced by the halo
calculation. This conclusion is strengthened by the fact that the
results for both elastic and inelastic scattering obtained by using
$4\hbar\omega$~\cite{Ka00} rather than $6\hbar\omega$ model space wave
functions and also by using the Paris Potential~\cite{La80} rather
than the Bonn-B interaction are very similar.  The validity of the
models used to predict the present data is corroborated by the very
good agreement obtained between the result of the $4\hbar\omega$ model
(406~mb) and the measured ($426 \pm 21$~mb) \cite{De00} reaction cross
section.

In conclusion, we have presented data for the elastic and inelastic
($2^+$) scattering of $^6$He from hydrogen at $40.9A$~MeV over a large
angular domain ($10^{\circ}$ to $80^{\circ}$). An excellent prediction
of both elastic and inelastic data has been made using a fully
microscopic, complex, non-local optical potential based on large basis
shell-model calculations of $^6$He with the incorporation of a neutron
halo. On the other hand, we have shown that the 2$^+$ state scattering
data are not reproduced by using the unaltered shell model wave
functions which overpredict the binding energy of the valence neutrons
and thus do not allow the halo to be formed.  The sensitivity of the
inelastic scattering data to the structure of $^6$He and the success
of the coordinate space scattering theories, based upon effective $NN$
interactions used successfully in analyses of proton scattering from
stable nuclei, open large perspectives for the study of the
microscopic structure of exotic systems.

%
%

\end{document}